\documentclass[aps,prb,twocolumn,showpacs]{revtex4}
\usepackage{graphicx}

\begin{document}

\title{Spin-Charge Coupling in lightly doped Nd$_{2-x}$Ce$_{x}$CuO$_4$}

\author{Shiliang Li}
\affiliation{
Department of Physics and Astronomy, The University of Tennessee, Knoxville, Tennessee 37996-1200, USA
}
\author{Stephen D. Wilson}
\affiliation{
Department of Physics and Astronomy, The University of Tennessee, Knoxville, Tennessee 37996-1200, USA
}
\author{David Mandrus}
\affiliation{
Condensed Matter Sciences Division, Oak Ridge National Laboratory, Oak Ridge, Tennessee 37831, USA}
\affiliation{
Department of Physics and Astronomy, The University of Tennessee, Knoxville, Tennessee 37996-1200, USA
}
\author{Bairu Zhao}
\affiliation{
Institute of Physics, Chinese Academy of Sciences, P.O. Box 603, Beijing 100080, China }
\author{Y. Onose}
\affiliation{
Spin Superstructure Project, ERATO, Japan Science and Technology, Tsukuba 305-8562, Japan}
\author{Y. Tokura}
\affiliation{
Spin Superstructure Project, ERATO, Japan Science and Technology, Tsukuba 305-8562, Japan}
\affiliation{
Correlated Electron Research Center,  Tsukuba 305-8562 Japan and Department of Applied Physics, University of Tokyo, Tokyo 113-8656, Japan}
\author{Pengcheng Dai}
\email{daip@ornl.gov}
\affiliation{
Department of Physics and Astronomy, The University of Tennessee, Knoxville, Tennessee 37996-1200, USA
}
\affiliation{
Condensed Matter Sciences Division, Oak Ridge National Laboratory, Oak Ridge, Tennessee 37831, USA}

\begin{abstract}
We use neutron scattering to study the influence of a magnetic field on
spin structures of Nd$_2$CuO$_4$. On cooling from room temperature, 
Nd$_2$CuO$_4$ goes through a series of antiferromagnetic (AF) phase transitions with 
different noncollinear spin structures.  
While a $c$-axis aligned magnetic field does not alter the basic zero-field noncollinear spin
structures, a field parallel to the CuO$_2$ plane can transform the noncollinear structure
to a collinear one (``spin-flop" transition), induce magnetic disorder along the $c$-axis, and cause 
 hysteresis in 
the AF phase transitions. By comparing these results directly to the magnetoresistance (MR) measurements of 
Nd$_{1.975}$Ce$_{0.025}$CuO$_4$, which has essentially the same AF structures as Nd$_2$CuO$_4$,
we find that a magnetic-field-induced spin-flop transition, AF phase hysteresis, and  spin $c$-axis disorder all affect the transport properties of the material. Our results thus provide direct evidence for the existence of a strong 
spin-charge coupling in electron-doped copper oxides.
\end{abstract}


\maketitle

\section{Introduction}
Understanding the role of magnetism in the transport properties and superconductivity
 of high-transition-temperature (high-$T_c$) copper oxides remains one of the important unresolved problems in the physics of
 transition metal oxides \cite{imada}. The parent compounds of high-$T_c$ cuprates are antiferromagnetic (AF) ordered Mott insulators composed 
of two-dimensional (2D) CuO$_2$ planes.  When holes or electrons are doped into these planes, the long-range AF ordered phase
is destroyed and the copper oxide materials become metallic and superconducting with persistent short-range AF 
spin correlations \cite{Kastner,daiprb,stock}. 
While much work over the past decade has focused on the interplay between magnetism and superconductivity because
spin fluctuations may mediate electron pairing for superconductivity \cite{hayden,tranquada},
understanding the relationship between AF order and transport properties through
the metal-insulator transitions (MIT) in these doped copper oxides is 
interesting in its own right.  

For instance, the parent compounds of hole-doped cuprates have collinear AF spin structure,
where each Cu$^{2+}$ spin is aligned opposite to its neighbors \cite{Kastner}. For La$_2$CuO$_4$ and 
La$_{2-x}$Sr$_x$CuO$_4$ in the lightly doped region, the Cu$^{2+}$ spins in the CuO$_2$ planes are 
slightly canted from the direction of the staggered magnetization to 
form a weak ferromagnetic (FM) moment \cite{thio1,thio2,lavrov01}. As a consequence, an
applied external magnetic field can manipulate the AF domain structure and induce large 
anisotropic magnetoresistance (MR) effect \cite{ando03}. In the case of the parent compounds of 
electron-doped materials such as Nd$_2$CuO$_4$ and Pr$_2$CuO$_4$,
the magnetic structures are noncollinear, where
spins in adjacent CuO$_2$ layers are 90 degrees ($90^\circ$) from each other [Fig. 1(c)],
 due to the pseudo-dipolar interaction between the rare-earth 
 (Nd$^{3+}$ and Pr$^{3+}$) and Cu$^{2+}$ ions \cite{lynn,ravi,petitgrand}.  Application of a magnetic field in the CuO$_2$ planes will induce 
a ``spin-flop'' transition by transforming the noncollinear structure
to a collinear one \cite{skanthakumarjap,skanthakumarprb,sumarlin}, and 
the critical field (${\bf B}_{SF}$) depends on the direction of the magnetic field with the
field along the Cu-Cu  (${\bf B}$$||[{\bar 1}10]$) 
direction generally has a smaller ${\bf B}_{SF}$ \cite{plakhty}.  A $c$-axis aligned magnetic field has no effect on the noncollinear spin structure \cite{masato}.

Recently, Lavrov {\it et al.} \cite{lavrov} have reported that an in-plane magnetic field can induce a large 
MR effect in lightly electron-doped copper oxide 
Pr$_{1.3-x}$La$_{0.7}$Ce$_x$CuO$_4$ ($x=0.01$). The authors find 
four-fold-symmetric angular-dependent MR oscillations for 
Pr$_{1.29}$La$_{0.7}$Ce$_{0.01}$CuO$_4$ in the low-temperature nonmetallic regime.
Similar data have been also obtained in nonsuperconducting Pr$_{1.85}$Ce$_{0.15}$CuO$_4$ independently \cite{fournier}. 
Since Pr$_{1.29}$La$_{0.7}$Ce$_{0.01}$CuO$_4$ has a noncollinear spin structure
at low temperatures and the critical fields for spin-flop transition and MR effects are similar, 
the novel MR phenomenon in this material has been attributed to the spin structure rearrangement from the noncollinear to the collinear state \cite{lavrov}. For a $c$-axis aligned magnetic field, the observed negative MR effect in 
the normal state of several different electron-doped cuprates has been interpreted as a result of 
two dimensional weak localization by disorder \cite{fournier00}, Kondo scattering from
Cu$^{2+}$ spins in the CuO$_2$ plane \cite{sekitani}, 
or spin scattering from field-induced magnetic 
droplets formed around impurities \cite{dagan1}.

While these recent MR measurements 
on Pr$_{1.29}$La$_{0.7}$Ce$_{0.01}$CuO$_4$ and Pr$_{1.85}$Ce$_{0.15}$CuO$_4$ clearly  
suggest a close coupling between spin-flop transition and MR effects \cite{lavrov}, 
the data may also be interpreted as partial rearrangement of magnetic domain walls 
by magnetic field to allow conductivity of electrons along a preferred direction \cite{fournier}.
In the latter case, the magnetic domain walls are segregated by 
the doped charge carriers into inhomogeneous patterns, such as stripes \cite{kivelson}. If the transport
properties in lightly electron-doped cuprates are indeed determined by spin reorientations and not 
by stripes, one would 
expect intimate correlations between the MR effects and spin-flop transitions 
in other families of electron-doped materials. Since 
Nd$_2$CuO$_4$ exhibits three AF phase transitions
with different noncollinear spin structures on cooling from room temperature \cite{skanthakumarjap,skanthakumarprb,masato},
far more complicated than the single AF phase transition found 
in Pr$_{1.29}$La$_{0.7}$Ce$_{0.01}$CuO$_4$ \cite{lavrov} or Pr$_{1.85}$Ce$_{0.15}$CuO$_4$ \cite{fournier}, 
a combined neutron scattering and MR investigation should shed new light on the 
interplay between spin and charge coupling in the material.

In this article, we describe our neutron scattering and MR measurements on single crystals of 
Nd$_2$CuO$_4$ and lightly electron-doped 
Nd$_{1.975}$Ce$_{0.025}$CuO$_4$, respectively. For neutron scattering, 
we choose to study Nd$_2$CuO$_4$ because of its complicated AF phase 
transitions \cite{skanthakumarjap}.
While previous work showed that a magnetic field applied parallel to  
the CuO$_2$ planes transforms the spins from 
the noncollinear to collinear AF structure \cite{skanthakumarjap,skanthakumarprb}, there is no systematic
work on how the field-induced collinear spin structure affects the zero-field AF 
phase transitions. We find that application of a ${\bf B}$$||[{\bar 1}10]$ field can induce $c$-axis spin disorder 
and hysteresis in the AF phase transitions. Since lightly electron-doping the insulating 
Nd$_2$CuO$_4$ induces enough charge 
carriers to allow transport measurements \cite{lavrov}
but does not change its basic AF spin structures \cite{yamada99}, we 
compare the MR effects in
Nd$_{1.975}$Ce$_{0.025}$CuO$_4$ to the neutron scattering results on Nd$_2$CuO$_4$. 
Surprisingly, we find that the transport properties of Nd$_{1.975}$Ce$_{0.025}$CuO$_4$ are very sensitive
to the modifications of spin structures in the system. Our results thus provide further evidence 
for the existence of a strong spin-charge coupling in electron-doped copper oxides. The organization of this
article is as follows. In Section II, we describe the experimental setup for neutron scattering and
transport measurements. Our neutron scattering results are presented in Section III while MR transport data
are shown in Section IV. In Section V, we compare the neutron scattering and transport data. Finally, Section VI 
summarizes the conclusions of our work.

\begin{figure}
\includegraphics[scale=.45]{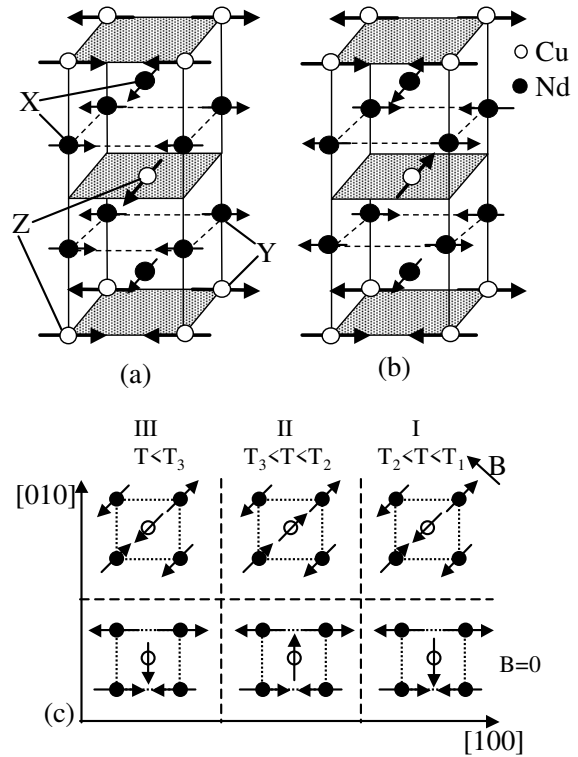}
\caption{Nd$_2$CuO$_4$ spin structures in (a) type-I/III and (b) type-II noncollinear states, 
where spins are indicated by the arrows. $X$, $Y$ and $Z$ represent the interactions between Nd-Nd, Nd-Cu and Cu-Cu spins respectively, as defined by Sachidanandam {\it et al.}\cite{ravi}. (c) the schematic phase diagram of CuO$_2$ planes in different phases at zero-field (bottom row) and ${\bf B}_{SF}||[{\bar 1}10]$ (top row). 
Here only Cu spins are shown for clarity. The filled and unfilled circles represent $L=0$ and 1/2 layers of Cu atoms respectively.
The $T_1$, $T_2$, and $T_3$ represent transition temperatures for the three different noncollinear phases at zero field \cite{lynn}. }
\end{figure}

\section{Experimental Setup}

We grew single crystals of Nd$_2$CuO$_4$ and Nd$_{1.975}$Ce$_{0.025}$CuO$_4$ by the traveling solvent floating-zone method. 
The samples were grown at a speed of 1 mm/hour under 4 atm O$_2$ pressure in a sealed quartz tube \cite{onose}. 
All the crystals are single domain as confirmed  
by a polarizing light microscope and Laue X-ray diffraction.  
The Nd$_2$CuO$_4$ single crystals are cylindrical and have dimensions of about 4 mm in diameter and 15 mm in length.  

The neutron scattering measurements on Nd$_2$CuO$_4$ were performed on the HB-1 and HB-3 triple-axis spectrometers 
at the high-flux-isotope reactor (HFIR), Oak Ridge National Laboratory (ORNL).  
We specify the momentum transfer $(q_x,q_y,q_z)$ in units of \AA$^{-1}$ as $(H,K,L)=(q_xa/2\pi,q_yb/2\pi,q_zc/2\pi)$ in reciprocal lattice units (r.l.u.). 
The lattice parameters of the tetragonal unit cells of Nd$_2$CuO$_4$ are $a=b=3.944$ {\AA} and $c=12.169$ {\AA}.
To prevent the samples from rotating under the influence of a magnetic field, they were clamped on solid aluminum 
brackets and placed inside a 7-T vertical field magnet \cite{masato}.
For the experiment, we use pyrolytic graphite as monochromator, analyzer and filters.  The collimations were, 
proceeding from the reactor to the detector, 48$^\prime$-40$^\prime$-sample-40$^\prime$-120$^\prime$ (full width at
half maximum or FWHM), and the final neutron energy was fixed at $E_f=14.78$ meV. The experiments were performed in the
$[H,H,L]$ scattering plane where the applied vertical field is along the $[{\bar 1}10]$ direction (${\bf B}$$||[{\bar 1}10]$)
in the CuO$_2$ plane.

\begin{figure}
\includegraphics[scale=.35]{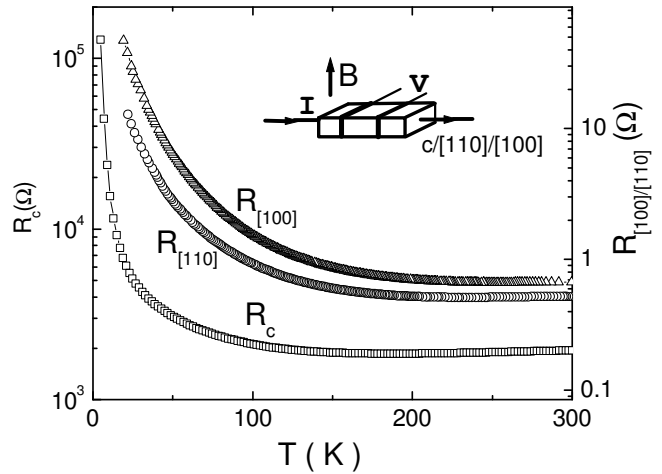}
\label{Fig2}
\caption{The temperature dependence of $R_c$, $R_{[100]}$ and $R_{[110]}$ at zero field, where the subscripts represent the current directions. The resistances in all three directions 
go up rapidly with decreasing temperature. The inset shows the regular four-points setup. 
Throughout the measurement, the $c$-axis of the crystal 
is always along the axis of rotation and the magnetic field rotates within the CuO$_2$ ($a$$b$) plane.}
\end{figure}

For our transport studies, we align and cut one Nd$_{1.975}$Ce$_{0.025}$CuO$_4$ 
single crystal into three rectangular blocks, whose
typical size was 3$\times$1$\times$0.5 mm$^3$.  Using the regular
four-points method, as shown in the inset of Fig. 2, the resistance was measured by the AC-transport option of a commercial 14-T physical property measurement system (PPMS). The AC currents were chosen along the [100], [110], and [001] directions
and the corresponding resistance were labeled as $R_{[100]}$, $R_{[110]}$, and $R_{c}$, respectively. 
Since Nd$_{1.975}$Ce$_{0.025}$CuO$_4$ has tetragonal crystal structure, the $a$- and $b$-axes 
are indistinguishable. As a consequence, a magnetic field along the $[100]$ ($[110]$) direction is equivalent to the $[010]$ ($[\bar{1}10]/[1\bar{1}0]$) direction. Figure 2 shows the temperature dependence of zero-field 
resistances in the three 
high symmetry directions. Similar to lightly doped Pr$_{1.29}$La$_{0.7}$Ce$_{0.01}$CuO$_4$ \cite{lavrov},
resistances in all directions of Nd$_{1.975}$Ce$_{0.025}$CuO$_4$ increase 
with decreasing temperature and show an insulating behavior at low temperatures. 
In addition, the resistivity data show no indication of the influence of  
AF phase transitions.  Since Nd$_{1.975}$Ce$_{0.025}$CuO$_4$ is only slightly doped away from
Nd$_2$CuO$_4$, it is reasonable to assume that these two systems have similar spin structures and AF
phase transitions \cite{yamada99}.

\section{Neutron Scattering Results}

Before describing our results on the influence of an in-plane magnetic field
on the magnetic order of Nd$_2$CuO$_4$, we briefly review its zero-field spin structures   \cite{skanthakumarjap,skanthakumarprb}.
When Nd$_2$CuO$_4$ is cooled from room temperature to $T_{1}\approx 275$ K, its Cu spins first order into the noncollinear type-I spin structure of Figs. 1 (a) and 1(c). On further cooling to $T_{2}\approx 75$ K, the Cu spins in the
adjacent layer rotate by 180$^\circ$ about the $c$-axis from the type-I phase 
and reorient the system into type-II phase [Fig. 1(b)]. Finally below $T_{3}\approx 30$ K, the Cu spins rotate back to their
original direction to form the type-III phase [see Fig. 1(c)] \cite{masato}.
Magnetic structure factors ($|F(1/2,1/2,L)|$) for noncollinear type-I and III phases 
at $(1/2,1/2,L)$ positions are
\begin{eqnarray}
&|F(L={\rm odd})|^2 = & 32(\gamma e^2/2mc^2)^2 \nonumber \\
&               &|f_{Cu}M_{Cu}+2\cos(2\pi Lz)f_{Nd}M_{Nd}|^2;    \nonumber \\
& {|F(L={\rm even})|^2} =& 32(\gamma e^2/2mc^2)^2 \nonumber\\
&                      & (2aL/c)^2/(2+(2aL/c)^2) \nonumber \\
&                      &|f_{Cu}M_{Cu}+2\cos(2\pi Lz)f_{Nd}M_{Nd}|^2.    \nonumber \\ 
\end{eqnarray}
For type-II noncollinear spin structure, we have
\begin{eqnarray}
&|F(L={\rm odd})|^2 = & 32(\gamma e^2/2mc^2)^2 \nonumber \\
&                   &(2aL/c)^2/(2+(2aL/c)^2) \nonumber \\
&               &|f_{Cu}M_{Cu}+2\cos(2\pi Lz)f_{Nd}M_{Nd}|^2;    \nonumber \\
& {|F(L={\rm even})|^2} =&    32(\gamma e^2/2mc^2)^2 \nonumber\\
&                      &|f_{Cu}M_{Cu}+2\cos(2\pi Lz)f_{Nd}M_{Nd}|^2;    \nonumber \\ 
\end{eqnarray}
where $\gamma e^2/2mc^2=0.2695\times 10^{-12}$ cm, $z=0.35$, $f_{Cu}$, $f_{Nd}$, $M_{Cu}$,
and $M_{Cu}$ are magnetic form factors and ordered magnetic moments for Cu and Nd ions,
respectively. 

From the magnetic structure factor calculations,
we find that the intensities of AF Bragg peaks at the $(1/2,1/2,L)$ positions
depend sensitively on the detailed spins arrangement.
For example, when Nd$_2$CuO$_4$ is cooled below the spins reorientation transition temperature $T_2$ [Fig. 1(c)], 
intensity of $(1/2,1/2,1)$ peak decreases while that of $(1/2,1/2,2)$ increases  \cite{lynn}.
On further cooling to below $T_3$, the scattering at these positions recover to their high temperature 
values [Figs. 3(a) and 3(b)].  Gaussian fits to the data show that the scattering is resolution limited with
the FWHM of 0.02 r.l.u for $(1/2,1/2,1)$ (or $\Delta L=0.02$ r.l.u.). In principle, one should calculate the 
coherence length of a Bragg peak from the formula of $N$-slit grating diffraction \cite{warren}. However, the
lineshape of our observed  diffraction
peaks is well described by a Gaussian, equivalent to the $N$-slit function in the limit of
large $N$.  By Fourier transform of the Gaussian peak, we estimate a minimum spin-spin coherence length
of $\sim$530 \AA\ using $CL=[4\ln(2)/\pi](c/\Delta L)$.
 
When a 5-T magnetic field is applied
along the $[{\bar 1},1,0]$ direction,
the noncollinear spin structures at different temperatures are 
transformed into collinear spin structures [Fig. 1(c)]. In phases I and III, $(1/2,1/2,L={\rm odd})$ peaks vanish
while $(1/2,1/2,L={\rm even})$ are enhanced with magnetic structure factors as
\begin{eqnarray}
& {|F_{c}(L={\rm even})|^2} =&    64(\gamma e^2/2mc^2)^2 \nonumber\\
&                            & (2aL/c)^2/(2+(2aL/c)^2) \nonumber\\
&                        &|f_{Cu}M_{Cu}+2\cos(2\pi Lz)f_{Nd}M_{Nd}|^2;    \nonumber \\ 
\end{eqnarray} 
For phase II, $(1/2,1/2,L={\rm even})$ peaks vanish
while $(1/2,1/2,L={\rm odd})$ reflections change as
\begin{eqnarray}
& {|F_{c}(L={\rm odd})|^2} =&    64(\gamma e^2/2mc^2)^2 \nonumber\\
&                        &|f_{Cu}M_{Cu}+2\cos(2\pi Lz)f_{Nd}M_{Nd}|^2.    \nonumber \\ 
\end{eqnarray} 
Figures 3(c) and 3(d) confirm that 
$(1/2,1/2,1)$ is enhanced while $(1/2,1/2,2)$ vanishes after spin-flop transition.
Figure 3(e) shows the integrated intensities at the $(1/2,1/2,1)$ and $(1/2,1/2,2)$ positions 
as a function of increasing temperature after 
a 5-T magnetic field is applied at 15 K to induce the 
type-III collinear state [Fig. 1(c)]. 
From the temperature dependence of their intensities, it is clear that 
spin-flop transitions occur between the three collinear states at similar temperatures
as that of the zero-field AF phase transitions \cite{masato}.

\begin{figure}
\includegraphics[scale=0.4]{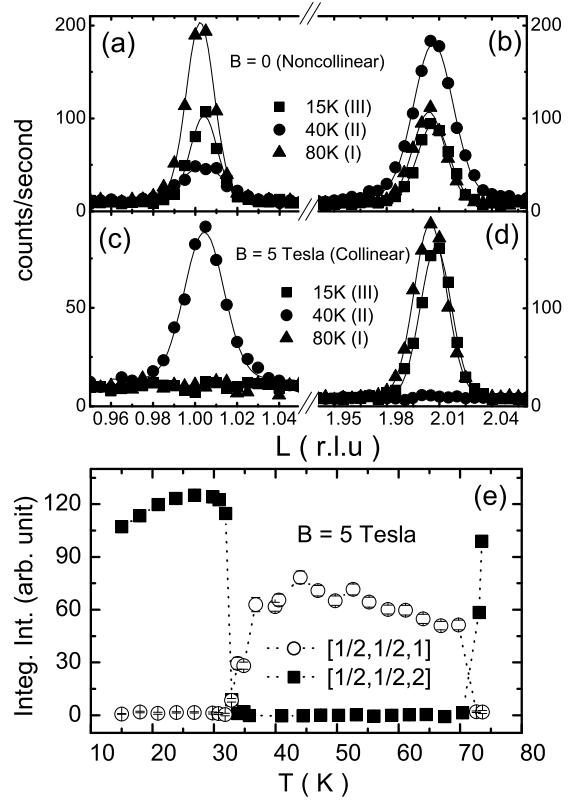}
\caption{Typical neutron scattering results around $(1/2,1/2,1)$ and $(1/2,1/2,2)$ Bragg peaks 
in the (a)(b) ${\bf B} = 0$ spin noncollinear and (c)(d) ${\bf B} = 5$ Tesla spin collinear states. 
We probed the magnetic Bragg peaks at 15 K, 40 K, and 80 K in type III, II, and I phases, respectively.
All peaks are fit by Gaussians on sloped backgrounds as shown by the solid lines. (e) The temperature 
dependence of the integrated intensities at $(1/2,1/2,1)$ and $(1/2,1/2,2)$ positions at ${\bf B} = 5$ Tesla. 
The type-III to II and type-II to I collinear phase transitions are around 30 K and 70 K, respectively.}
\end{figure}

\begin{figure}
\includegraphics[scale=.4]{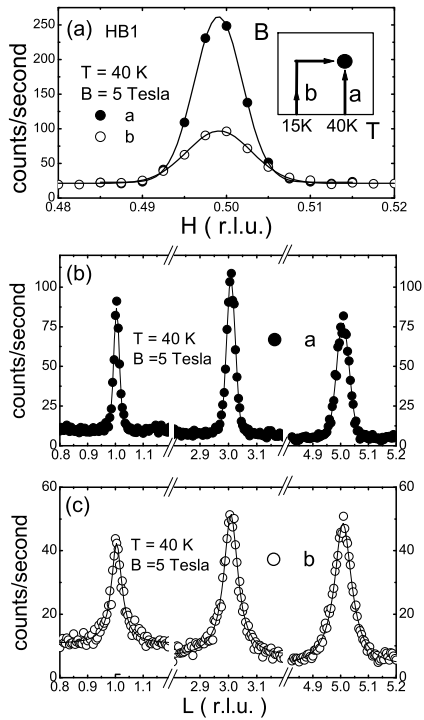}
\caption{(a) The in-plane $[H,H,1]$ scans across $(1/2,1/2,1)$ in two different processes at 40 K.
The schematic diagrams of processes \textbf{a} and \textbf{b} are shown in the inset. (b)
The long $[1/2,1/2,L]$ scan in process \textbf{a}, where $L$ changes from 0.8 to 5.2 r.l.u. 
Only three resolution-limited peaks are found around $L = 1$, 3 and 5. 
(c) Same scans as (b), but through process \textbf{b}. Note that the $L$-widths of these peaks
become considerably broader.
The temperature and field in (b) are identical 
as that of (c), \textit{i.e.} $T = 40$ K and ${\bf B} = 5$ Tesla. 
However, the processes of applying field are different, as shown in the inset of (a). 
The peaks in (b) and (c) are fitted by Gaussians and Lorentzians, respectively.}
\end{figure}

To test whether the field-induced type-II collinear spin structure
depends on the magnetic field hysteresis, we perform neutron 
experiments at 40 K in two ways. We first cool the
sample at zero-field to 40 K and then increase the ${\bf B}$$||[{\bar 1}10]$ field to 5-T as sketched in
the inset of Fig. 4(a) (process {\bf a}).
The $[1/2,1/2,L]$ scan from $0.8<L<5.2$ r.l.u. shows resolution-limited peaks 
around $L = 1, 3$ and 5 as shown in Fig. 4(b). Since the $(1/2,1/2,1)$ peak has a Gaussian lineshape 
with FWHM of 0.02 r.l.u. along the $L$ direction, we estimate that the $c$-axis magnetic 
coherence length is around 530 \AA.
Now, if we applied the 5-T field at 15 K and then increased the temperature to 40 K (process {\bf b}), 
the $(1/2,1/2,L)$ ($L=$ odd) peaks remain but with much different lineshape [Fig. 4(c)]. 
Instead of having resolution-limited Gaussian lineshape, the peaks 
are Lorentzians (The Ornstein-Zernike form) with much broader widths but the same integrated intensity as that in Fig. 4(b). 
For example, while the FWHM of the $(1/2,1/2,1)$ peak increases to 0.046 r.l.u. from 
0.02, its peak intensity also drops by $\sim$50\% [Figs. 4(b) and 4(c)]. These observations 
suggest that the $c$-axis spin-spin correlation function decays
exponentially with a much shorter coherence length [$\sim$84 \AA\ using $1/\kappa$ where $\kappa$ is
HWHM in \AA$^{-1}$ of the Lorentzian in Fig. 4(c)] in process \textbf{b}. 
Since the only difference between the type-III and II collinear states is 
the 180$^\circ$ spins rotation in adjacent layers [Fig. 1(c)], the short spin-spin 
coherence length in Fig. 4(c) suggest the presence of a $c$-axis spin disorder. On the other hand,
in-plane scans along the $[H,H,L]$ ($L=1$, 3, 5) direction only show slight broadening when
field is applied at low temperature in process \textbf{b} [Fig. 4(a)], thus suggesting most of the spin disorder 
occurs along the $c$-axis. 

To further investigate how hysteresis in application of a magnetic field
can affect the spin arrangements in Nd$_2$CuO$_4$, 
we studied its $c$-axis coherence lengths in all three collinear phases shown in Fig. 1(c). 
There are three different ways to reach the expected temperature and field of 40 K and 5-T in the type-II collinear spin phase,
 which are labeled as \textbf{a}, \textbf{b} and \textbf{c} in the inset of Fig. 5(a). 
Precesses \textbf{a} and \textbf{b} are the same as Fig. 4 and
process \textbf{c} involves getting to 80 K in zero field, applying the 5-T field, and then cooling the sample to 40 K.
In other words, the spin system changes from the noncollinear state to the collinear state and remains in type-II
collinear spin phase during the process \textbf{a}, whereas the system undergoes phase transitions 
from type-III and type-I collinear states to type-II collinear state
in processes \textbf{b} and \textbf{c}, respectively [Fig. 1(c)].  

\begin{figure}
\includegraphics[scale=.4]{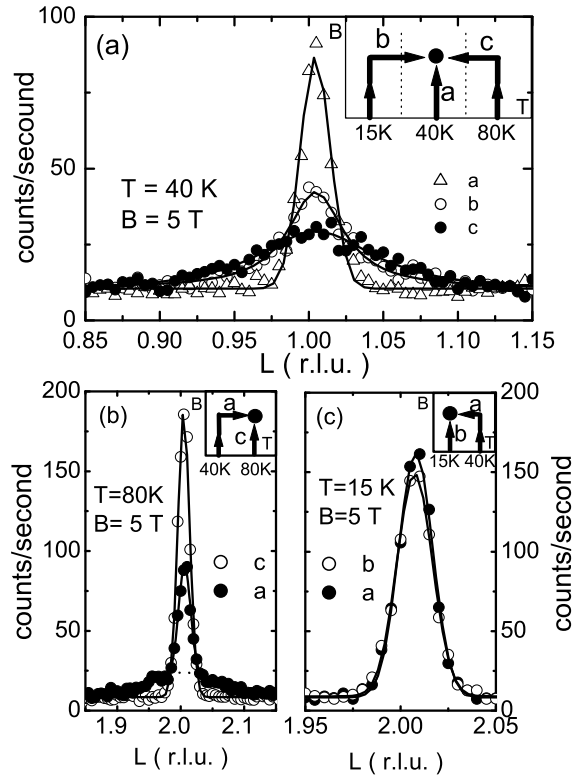}
\caption{Magnetic field hysteresis effects on AF spin structures of Nd$_2$CuO$_4$ in all three phases.
(a) $(1/2,1/2,1)$ Bragg position in phase II and (b),(c) $(1/2,1/2,2)$ position in phase I and III, respetively.  
The corresponding processes of field and temperature 
are shown in the inset of each figure.}
\end{figure}

By comparing the $(1/2,1/2,1)$ data of Fig. 4 
in processes \textbf{a} and \textbf{b} with that of \textbf{c} in Fig. 5(a),
it becomes clear that processes \textbf{b} and \textbf{c} have 
considerably broader widths.
This means that \textbf{b} and \textbf{c}
processes induce large $c$-axis spins disorder and the amount of disorder depends on the history
of field application. When the system is in type-I phase at 80 K, the collinear spin phase can be
induced either by simply applying a field at 80 K [process \textbf{c} in the inset of Fig. 5(b)],
or by having a type-II collinear phase at 40 K and then increasing the temperature to 80 K
 [process \textbf{a} in the inset of Fig. 5(b)].
Clearly, the $[1/2,1/2,L]$ ($L=2$) scans in Fig. 5(b) show 
resolution-limited Gaussian peak in \textbf{c} but 
two components lineshape
with a sharp Gaussian peak on a broad Lorentzian background in process \textbf{a}. 
These results again suggest that magnetic field hysteresis affects mostly the spin arrangements
along the $c$-axis. Finally, when temperature is in
the type-III collinear phase at 15 K, we find no obvious difference between the two processes in
the inset of Fig. 5(c): one of which is applying field at 40 K and then decreasing the temperature 
to 15 K (process \textbf{a}), and the other is applying the field at 15 K directly (process \textbf{b}).
Therefore, spin disorder appears whenever phase transition occurs between two different
collinear states except for the transition from type-II to III.

Figure 6 shows the magnetic field dependence of the $(1/2,1/2,1)$ peak at $T = 40$ K under different conditions. 
When a ${\bf B}$$||[{\bar 1}10]$ field is applied from the zero-field noncollinear type-II state at 
40 K [process \textbf{a} in the inset of Fig. 6(a)], a noncollinear to collinear spin-flop transition 
occurs around 1-T, and the FWHM of the peak does not change during the process [Figs. 6(a) and 6(b)]. This suggests 
that the entire AF structure responds to the influence of the applied field.
If we warm to 40 K using process \textbf{b} shown in the inset of Fig. 6(a), 
the $(1/2,1/2,1)$ peak has a broad FWHM along the $c$-axis but with the same integrated intensity 
as process \textbf{a}. As a function of decreasing magnetic field at 40 K, the FWHM of 
$(1/2,1/2,L)$ at $L=1$ decreases continuously until reaching the value of process \textbf{a}.
This suggests that long-range $c$-axis spin coherence length is restored in the process.

\begin{figure}
\includegraphics[scale=.4]{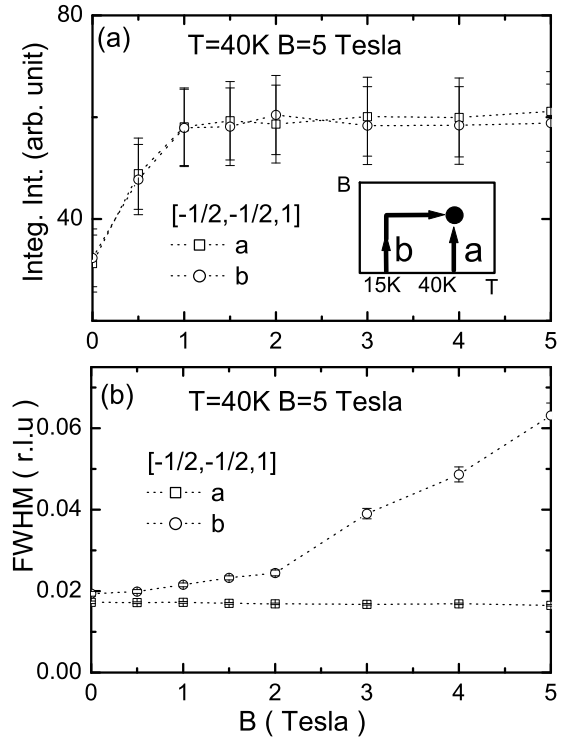}
\caption{ The field dependence of (a) integrated intensities and (b) FWHMs at the $(1/2,1/2,1)$ 
position during two different processes shown in the inset of (a). 
The error bars in (a) are obtained by taking the square root of total summed intensty in $L$-scans.}
\end{figure}

\begin{figure}
\includegraphics[scale=.4]{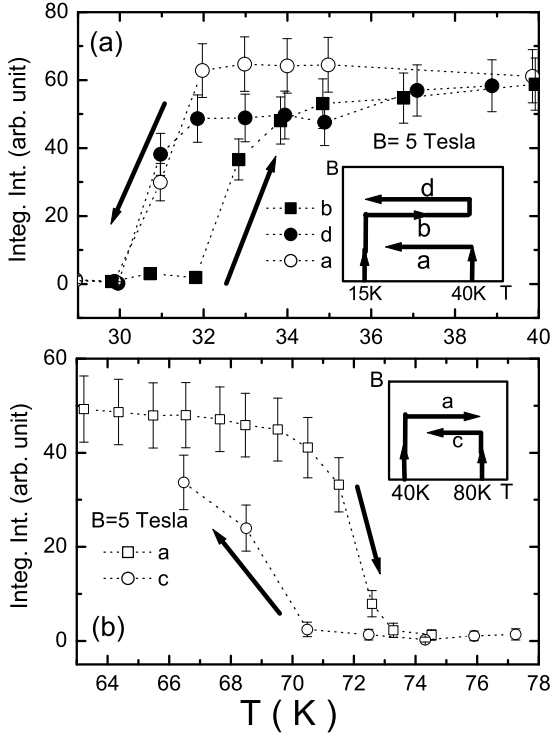}
\caption{ The integrated intensities of $(1/2,1/2,1)$, which shows temperature hysteresis 
across the transition (a) between type-II and III phases, (b) between type-I and II phases. 
The arrows in the figures indicate the direction of changing temperatures. 
The insets show the detailed processes of field-temperature hysteresis.}
\end{figure}

Because the spin disorder is related to the phase transitions under field, it is natural to ask what happens 
to the transition itself under different conditions. Here, we use 
two ways to pass the type-II to type-III phase transition, and label them as 
processes \textbf{a} and \textbf{d} in the inset of Fig. 7(a). We carefully monitor
the $(1/2,1/2,1)$ peak as the system passing the transition under different conditions.
At each measured temperature, we wait 10 to 15 minutes to ensure that the peak intensity 
has no time dependence and the outcome of the experiment is summarized in Fig. 7. 
Surprisingly, there is a clear hysteresis in the phase transition behavior, 
\textit{i.e.} the transition temperature of process \textbf{b} is about 2 K higher than that of process \textbf{a}.

To understand how this hysteresis occurs, we decreased the temperature following process \textbf{b}, 
\textit{i.e.} we probe the $(1/2,1/2,1)$ peak in process \textbf{d} after \textbf{b}. 
Figure 7(a) indicates that 
the transition temperatures of processes \textbf{a} and \textbf{d} are identical.
While this suggests that spin disorder itself may not induce the lower transition
temperature in decreasing temperature processes, the observed hysteresis must be 
associated with the field-induced collinear spin structures and their free energy differences
in different phases. Similar hysteresis behavior can also be found for the transition between collinear type-II and I phases
[Fig. 7(b)]. In this case, 
the integrated intensities of $(1/2,1/2,1)$ during process \textbf{c} are
 much smaller than those during process \textbf{a} below 70 K, while the intensities of $(1/2,1/2,2)$ 
remain zero thus ruling out a possible mixture of the two phases.

In previous work on Pr$_2$CuO$_4$ \cite{petigrand2}, which has a noncollinear 
type-I/III spin structure [Fig. 1(a)] \cite{sumarlin}, diffuse scattering
associated with interplane short-range order
was observed above the spin-flop transition critical field (3.1-T) around the
forbidden AF position $(1/2,1/2,1)$ at $T=1.5$ K. Since similar diffuse scattering
was not observed around $(1/2,1/2,2)$, the authors suggest that the diffuse scattering arises from
the persistent mid-range interplane correlations \cite{petigrand2}. For Nd$_2$CuO$_4$,
we find no evidence for similar short-range order spin order at $(1/2,1/2,1)$ (Figs. 2-5). 
Instead, we find clear evidence for  
field-induced AF phase transition hysteresis and spin disorder. If the spin degree of freedom is strongly coupled to the  
charge carriers, one would expect 
to observe changes in electrical transport properties uniquely associated with  
field-induced spin disorder and hysteresis in these materials. 
By performing systematic MR measurements in lightly-doped Nd$_{1.975}$Ce$_{0.025}$CuO$_4$,
which have essentially the same AF structure 
as Nd$_2$CuO$_4$ but with enough charge
carriers for resistance measurements, we can directly compare the transport data with neutron scattering
results. The following Section will describe such comparison.

\section{Magnetoresistance results}

\begin{figure}
\includegraphics[scale=.4]{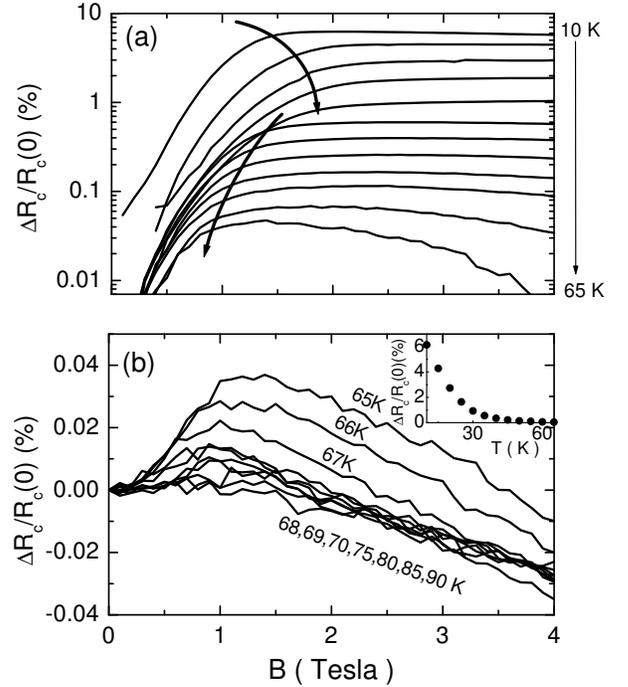}
\caption{ $\Delta R_c/R_c(0)$ as a function of increasing magnetic field at different temperatures 
around (a) 30 K and (b) 70K. Note that these temperatures are close to $T_3$ and $T_2$, respectively.  
The applied magnetic field is in the CuO$_2$ $ab$-plane along the $[{\bar 1}10]$ direction. 
The arrows in (a) indicate the temperature dependence of ${\bf B}_{SF}$. 
The inset in (b) plots the maximum of $\Delta R_c/R_c(0)$ as a function of increasing temperature.
It shows $1/T$ behavior. Note that the vertical axes are log scale in (a) and linear scale in (b). }
\end{figure}

In the work of Lavrov and co-workers on Pr$_{1.29}$La$_{0.7}$Ce$_{0.01}$CuO$_4$ 
\cite{lavrov}, the similarities between the critical field
for the spin-flop transition and the rapid increase of MR have been taken as evidence for 
spin-charge coupling. Because there exists three spin phases in 
Nd$_2$CuO$_4$ whereas only one in Pr$_{1.29}$La$_{0.7}$Ce$_{0.01}$CuO$_4$, one would expect some new phenomena related to those phases and the transitions between them. Figure 8 shows the 
$c$-axis MR effect $\Delta R_c/R_c(0)$ of Nd$_{1.975}$Ce$_{0.025}$CuO$_4$ at different temperatures. 
As a function of increasing field along the 
$[{\bar 1} 1 0]$ direction, $R_c/R_c(0)$ initially increases quickly but then descends slightly. 
The increase in the critical field for spin-flop transition, ${\bf B}_{SF}$, with increasing temperature
for $T$ below 30 K is consistent with earlier neutron scattering experiments on Nd$_2$CuO$_4$ \cite{skanthakumarjap}.
The increase of ${\bf B}_{SF}$ with increasing temperature ends above 
$T_3=31$ K, where the type-III to type-II spin-flop transition occurs [Fig. 8(a)].
Compared with the results on Pr$_{1.29}$La$_{0.7}$Ce$_{0.01}$CuO$_4$ \cite{lavrov}, 
the data suggest that the changes in MR effect originate from the
differences of ${\bf B}_{SF}$ in three different spin phases. 
Below 30 K, ${\bf B}_{SF}$ increases slightly with increasing temperature.
It then decreases with increasing temperature beyond 30 K. 
Finally, when the system changes to type-I phase above 67 K, the sharp increase of $R_c/R_c(0)$ at low fields almost disappears, and all $\Delta R_c/R_c(0)$ data nearly fall into one curve [Fig. 8(b)]. 
Similar phenomena are also found for $R_{[110]}/R_{[110]}(0)$ and $R_{[100]}/R_{[100]}(0)$. 
The temperature dependence of the maximum of $\Delta R_c/R_c(0)$ is shown 
in the inset of Fig. 8(b), which can be fit by the $1/T$ function similar to low temperature
intensity changes in Nd$_2$CuO$_4$ \cite{masato} and Nd$_{1.85}$Ce$_{0.15}$CuO$_4$ \cite{kang}. 

\begin{figure}
\includegraphics[scale=.4]{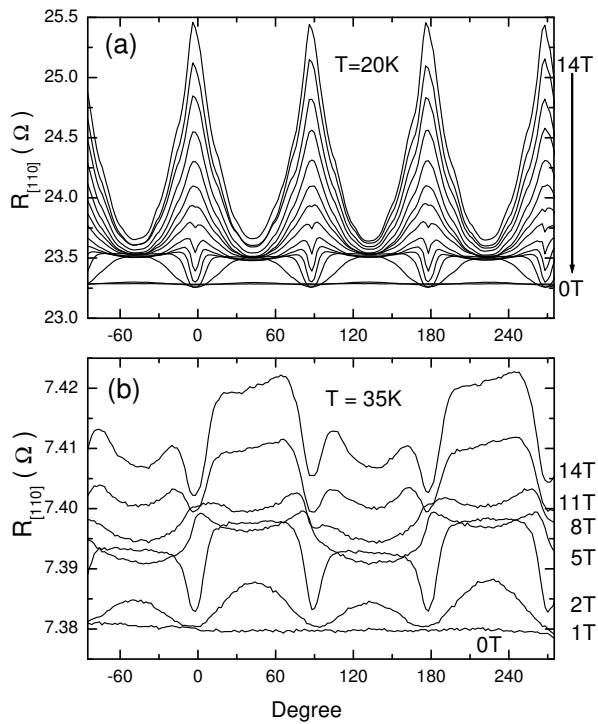}
\caption{Angular dependence of $R_{[110]}$ at (a) $T = 20$ K in type-III phase 
and (b) $T = 35$ K in type-II phase, where $[110]$ indicates the current direction.
The horizontal axes represent the angle (in degrees) 
between the magnetic field and the $a$- or $b$-axis which cannot be
distinguished for the tetragonal system. Note that data in (a) at 1-T data show non-observable MR effect while there are
clear MR effects at 1-T in (b).}
\end{figure}

In previous work, a fourfold angular oscillation in MR effect has been identified in 
Pr$_{1.29}$La$_{0.7}$Ce$_{0.01}$CuO$_4$ \cite{lavrov} and Pr$_{1.85}$Ce$_{0.15}$CuO$_4$ \cite{fournier} 
for an applied magnetic field in the CuO$_2$ plane. For lightly doped Nd$_{1.975}$Ce$_{0.025}$CuO$_4$,
we expect to observe similar fourfold oscillation behavior. In addition, we hope to determine whether 
transitions across different spin phases affect the MR. 
To study the angular dependence of resistance, we rotated the sample with respect to the $c$-axis 
which remained perpendicular to the magnetic field, \textit{i.e.} the field rotated within the $ab$-plane. 
The rotation angle is defined to be zero when ${\bf B}||[100]/[010]$. As shown in Fig. 9(a),  
a fourfold feature in $R_{[110]}$ similar to the earlier results is found as a function of the rotation angle.
While $R_{[110]}$ along the $[110]/[1\bar{1}0]$ direction is much higher than that along the 
$[100]/[010]$ direction at low fields, the situation reverses itself at high fields.

\begin{figure}
\includegraphics[scale=.4]{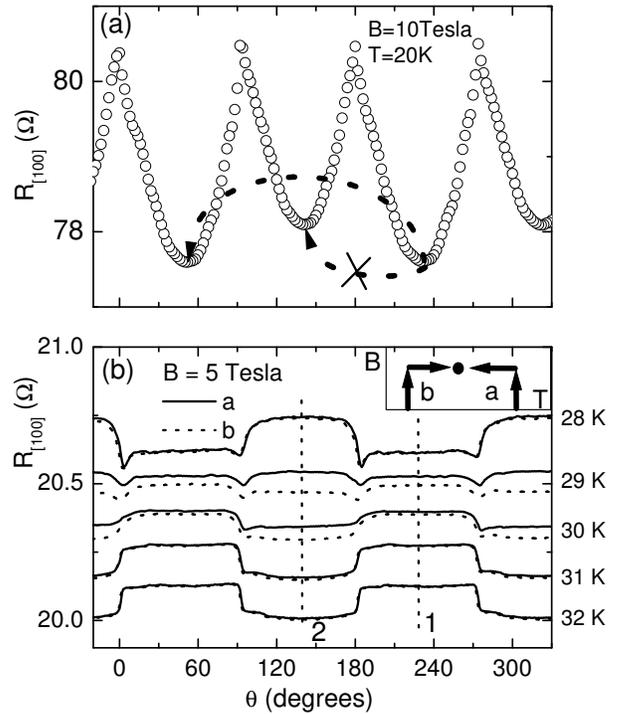}
\caption{(a) Angular dependence of $R_{[100]}$ at 20 K and 10-T. 
While we still observe the fourfold feature, the MR effects do not have the expected 90$^\circ$
symmetry. We speculate that this is due to the small misalignment of the sample with respect to the
applied field. (b) Hysteresis behavior of $R_{[100]}$ around the transition between the type-II and III phases
at 5-T. At each measured temperature, a constant has been subtracted from the resistance values in processes
\textbf{a} and \textbf{b} for clarity.
The solid and dotted lines represent increasing (\textbf{b}) and decreasing (\textbf{a}) 
temperatures, respectively, as shown in the inset.}
\end{figure}
 
Assuming a strong spin-charge coupling, one can understand the microscopic process
of the MR effect as follows. When an in-plane magnetic field is applied along 
the $[110]/[{\bar 1}10]$ direction, spins in the type-III noncollinear structure of Fig. 1(c)
rotate continuously to form the collinear structure perpendicular to the field \cite{skanthakumarjap}.
While these diagonal (Cu-Cu) directions are easy axes in the collinear spin structure with relatively
small ${\bf B}_{SF}[{\bar 1}10]$, a perfectly
aligned field along the $[100]/[010]$ direction induces a first-order spin-flop transition 
with a much larger critical field ${\bf B}_{SF}[100]$ \cite{plakhty}. 
For a magnetic field in the intermediate directions,
it first induces a transition into the collinear state and then smoothly rotates the spins to positions
perpendicular to the field \cite{lavrov,plakhty}. 
If an applied field is less than 1-T, there is no spin-flop transition at any 
field orientation and thus no MR effect, consistent with Figs. 8(a) and 9(a).
When an applied field (2-T $\leq {\bf B} \leq 6$-T) is larger than ${\bf B}_{SF}[{\bar 1}10]$ but
smaller than ${\bf B}_{SF}[100]$, one would expect to observe a spin-flop transition, and therefore the MR effect, for fields
along the $[110]/[\bar{1}10]$ direction but not along the $[100]/[010]$ direction.  This is exactly what we find 
in Fig. 9(a). Finally for ${\bf B} \geq {\bf B}_{SF}[100]$, the aligned collinear spins will simply follow the rotation
of the field in all directions. Here the magnitude of the MR effect for ${\bf B}||[100]/[010]$ 
is larger than that for ${\bf B}||[110]/[{\bar 1}10]$, therefore causing a new four-fold MR oscillation
with 90$^\circ$ angles shift from that in ${\bf B}_{SF}[{\bar 1}10]\leq {\bf B}\leq {\bf B}_{SF}[100]$.

\begin{figure}
\includegraphics[scale=0.4]{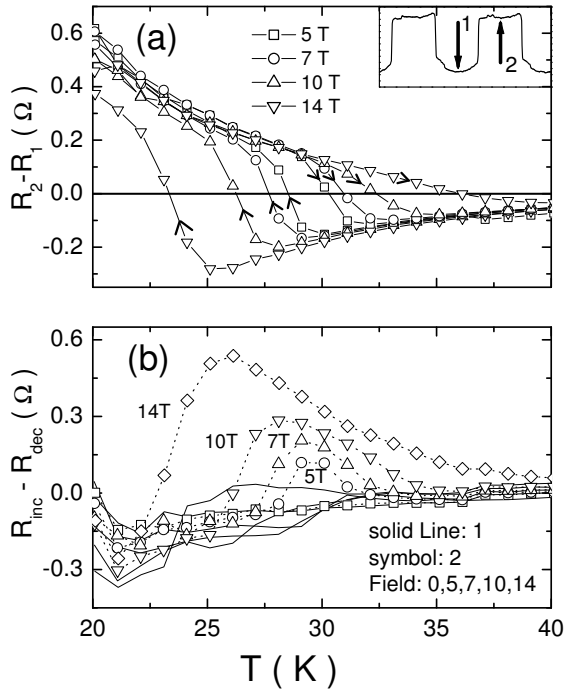}
\caption{(a) $R_2-R_1$ during increasing and decreasing temperature processes, as shown by the arrows. 
The positions 1 and 2 are defined in the inset of (a). With increasing field, the widths of hysteresis become larger. (b)$R_{inc}-R_{dec}$, the differences between increasing and decreasing temperature processes 
at positions 1 and 2 are plotted by lines and symbols, respectively.}
\end{figure}

As the temperature of the system is increased to 35 K in the type-II phase, the magnitude of ${\bf B}_{SF}[{\bar 1}10]$
becomes smaller than that at 20 K \cite{skanthakumarjap,skanthakumarprb,lavrov}. 
This explains the observed fourfold MR oscillations in Fig. 9(b) at 1-T, while similar oscillations are only 
seen for fields above 2-T at 20 K [Fig. 9(a)]. On increasing the applied field from 1-T,
the oscillations change from sinusoidal shape to a flattish (or concave)
 top but the symmetry of the oscillations as a function of rotational angle remains 
even up to the maximum field of 14-T. The data are clearly different from that at 20 K for fields
 above 6-T, where MR effects are maximal along the $[100]/[010]$ directions for ${\bf B}\geq 6$-T.
To consistently interpret the MR data of Figs. 9(a) and (b), we speculate that ${\bf B}_{SF}[100]$ is 
larger than the maximum applied field of 
14-T around 30 K. While this scenario is difficult 
to prove because neutron experiments in the ${\bf B}||[100]/[010]$ geometry have not
yet been carried out at this temperature \cite{plakhty},  
 a large ($\geq$ 4.5 meV) in-plane spin-wave gap associated with
the Nd$^{3+}$-Cu$^{2+}$ interactions has been reported in the type-II phase of Nd$_2$CuO$_4$ \cite{bourges93}.
If closing such a spin-wave gap is required to induce the spin-flop transition
in the ${\bf B}||[100]/[010]$ geometry, the critical field necessary 
to produce Zeeman energy larger than 4.5 meV will exceed 14-T assuming only Cu$^{2+}$ 
contributions [see equation (2) below].
Alternatively, one might imagine that the bigger 
low-temperature MR effect 
in the ${\bf B}||[100]/[010]$ direction is somehow related to the larger Nd$^{3+}$ moments and/or  
the first order nature of the
spin-flop transition in this direction \cite{lavrov,fournier}. While how MR is affected by the 
Nd$^{3+}$-Cu$^{2+}$ coupling is unknown,   
a small misalignment of the sample with respect to the magnet around 
$[100]/[010]$ directions can affect dramatically 
the observed MR.  Such misalignment may also explain the slightly different MR values 
at 45$^\circ$ and 135$^\circ$ in Fig. 9(b). We note that similar, but less obvious, 
behavior is also present in Fig. 9(a) and in previously reported MR data \cite{lavrov,fournier}.

Figure 10(a) shows the angular dependence of $R_{[100]}$ at 20 K and 10-T.
While $R_{[100]}$ has fourfold oscillations similar to that of $R_{[110]}$ at the same temperature [Fig. 9(a)],
the MR differences between 45$^\circ$ and 135$^\circ$ are more obvious. 
With increasing temperature, the fourfold oscillations are replaced by a square-wave-like feature [Fig. 10(b)]
similar to Fig. 9(b). Note that our neutron scattering revealed a clear hysteresis through the type-III to II 
collinear phase transition [Fig. 7(a)]. 
To see if MR follow such hysteresis, we performed careful measurements on field-warming 
 and cooling as processes \textbf{b} and \textbf{a}, respectively [see the inset of Fig. 10(b)].
 The outcome in Fig. 10(b) shows clear hysteresis across the type-III to II transition, consistent with the neutron
 scattering results of Fig. 7. In addition, we find that the relative value of $R_{[100]}$ shifts 90$^\circ$ 
 across the transition, suggesting that the MR effects are sensitive to the differences in
 the type-III and II collinear spin structures [Fig. 1(c)]. 
 If we define the resistance at 135$^\circ$ position as
$R_2$ and 225$^\circ$ as
$R_1$, $R_2$ is larger than $R_1$ in the type-III collinear phase
while the reverse is true in the type-II collinear phase.

Using the resistance difference between positions 1 and 2, we probe the phase transition between
type-III and II collinear states in great detail.
If there is no field-induced hysteresis, $R_2-R_1$ should be the same for either warming or cooling.
Figure 11(a) indicates that this is not the case. On warming, the type-III collinear to II 
collinear transition temperature
increases with increasing field. On the contrary, the type-II to III transition temperature decreases
with increasing field on cooling.  As a consequence, the width of the hysteresis increases with increasing
field and can be as large as $\sim$15 K at 14-T [Fig. 11(a)].  At 5-T, 
the width of the hysteresis is about 2 to 3 K, completely consistent with
the neutron scattering results of Fig. 7. 
Figure 11(b) shows the differences in resistance at positions 1 and 2 
between increasing and decreasing temperature processes for various applied fields.
The results also suggest an increasing hysteresis in the phase transition with 
increasing magnetic fields, consistent with Fig. 11(a).

\begin{figure}
\includegraphics[scale=0.35]{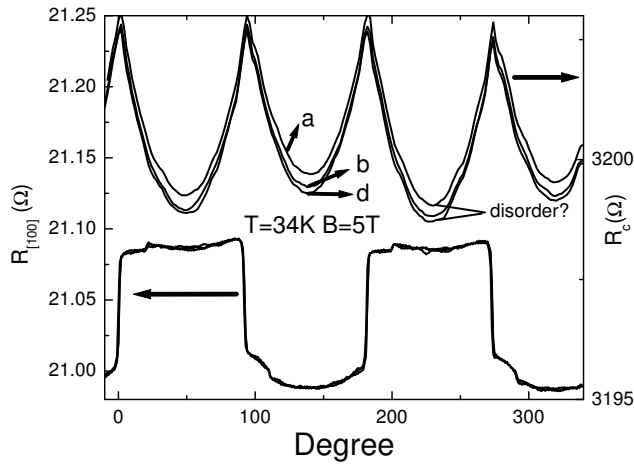}
\caption{Angular dependence of $R_{[100]}$ and $R_c$ at 34 K and 5-T through processes 
\textbf{a}, \textbf{b}, and \textbf{d} similar to that defined in Fig. 7(a),
except in this case the final temperature was fixed at 34 K. 
Process \textbf{a} does not across the phase transition temperature,
and therefore does not exhibit disorder. Processes \textbf{b} and \textbf{d}
have spin disorder.  While clear differences are seen in $R_c$, 
there are no observable differences in 
$R_{[100]}$ for these different processes.
}
\end{figure}

Finally, we describe the transport measurements associated with
the $c$-axis disorder seen by neutron scattering.
Following the same processes as the inset in Fig. 7(a), we have measured the resistance of Nd$_{1.975}$Ce$_{0.025}$CuO$_4$
in three different current directions, $R_{[100]}$, $R_{[110]}$, and $R_c$. 
For the in-plane resistances, 
we find no distinguishable difference between $R_{[100]}$ and $R_{[110]}$ after waiting one hour for each measurement. 
The results of $R_{[100]}$ at 34 K and 5-T in Fig. 12 show overlapping curves for processes 
of \textbf{a}, \textbf{b}, and \textbf{d}. On the other hand, $R_c$ displays clear distinctions  
among the varying processes as shown in Fig. 12. Since 
the resulting differences in $R_{[100]}$ for processes \textbf{a}, \textbf{b}, and \textbf{d}
are less than $5\times 10^{-5}$, we can safely conclude that 
the observed deviations in $R_c$ 
among these processes are intrinsic and may originate from the spin disorder along the $c$-axis. More
work is needed to understand precise relationship between spin disorder and charge
transport properties.


\section{Discussion}

Until now, the most successful theory to understand the spin properties of Nd$_2$CuO$_4$ 
 is based on pseudo-dipolar interaction (PDI) originally proposed by Van Vleck in 1937,
\begin{equation}
V_{pd}=\frac{1}{2}\sum_{\ell\ell^{'}}{V(\textbf{R}_{\ell\ell^{'}})(\textbf{S}_{\ell^{'}}\hat{\textbf{R}}_{\ell\ell^{'}})(\textbf{S}_{\ell}\hat{\textbf{R}}_{\ell\ell^{'}})},
\end{equation}
\noindent where $\ell$ and $\ell^{'}$ denote the lattice sites and the function $V(R)$ decreases faster than $R^{-3}$ as $R\rightarrow\infty$. To explain the reorientation of the spin structure, Sachidanandam \textit{et al.}\cite{ravi} considered three major interplane interactions between Nd-Nd, Nd-Cu and Cu-Cu, 
labeled as $X$, $Y$ and $Z$ respectively in Fig. 1(a). The interactions
$X$ and $Z$ tend to generate the type-III or I spin structures, and while 
$Y$ prefers type-II phase. 
The interactions between spins are proportional to their local susceptibilities ($m$).
Since $m_{Nd}$ is proportional to  $1/T$ \cite{ravi} and $m_{Cu}$ varies little below 40 K \cite{lynn},
 $X \propto 1/T^2$, $Y \propto 1/T$ and $Z \approx Constant$. 
With decreasing temperature, Cu-Cu ($Z$) interactions initially turn on below $T_1$ 
and Nd$_2$CuO$_4$ orders antiferromagnetically with the type-I spin structure [Fig. 1(a)]. On
cooling to intermediate temperature $T_2$, Nd-Cu ($Y$) interactions become important and 
the system transforms to the type-II noncollinear spin structure. Finally below $T_3$, Nd-Nd ($X$)
interactions dominate and induce the type-III noncollinear spin structure (Fig. 1).

The noncollinear spin structures of Nd$_2$CuO$_4$ have a small spin-wave anisotropy gap $\Delta_0$ 
at zero field \cite{petitgrand}. When an in-plane field is applied, the Zeeman energy 
shifts the spin-wave dispersion and closes the anisotropy energy gap, resulting 
a transition from noncollinear to collinear
spin-flop phase. Petitgrand \textit{et al.} \cite{petitgrand} have given 
the critical field of spin-flop transition when the field is along $[{\bar 1}10]$ direction, 
\begin{equation}
B_{SF}[{\bar 1}10]=\frac{\Delta_0}{gm\mu_B},
\label{hsf}
\end{equation}
\noindent where $\Delta_0$ is in-plane spin-wave gap at zero field, $g$ is Landau factor, 
$m$ the effective moment, and $\mu_B$ the Bohr magneton. This equation has been successfully used to explain the temperature dependence of ${\bf B}_{SF}[{\bar 1}10]$ for Pr$_2$CuO$_4$ \cite{sumarlin}. For Nd$_2$CuO$_4$, the in-plane Cu spin-wave gap has a $1/T$ dependence  and the out-of-plane gap is essentially temperature independent \cite{ivanov}. 
In addition, Nd spin-waves exhibit anisotropic gaps at low temperatures \cite{pyka}. Since an applied field of a
few Tesla in the CuO$_2$ will not change the large ($>5$ meV) Cu spin-wave gap in type-III phase below 30 K \cite{bourges93}, 
the spin-flop transition there is most likely induced by closing the Nd spin-wave gap.

The existence of a spin-charge coupling has been suggested in lightly electron-doped 
Pr$_{1.29}$La$_{0.7}$Ce$_{0.01}$CuO$_4$, but a detailed microscopic understanding of
how such coupling occurs is still lacking \cite{lavrov}. 
If the itinerant electrons are coupled to the localized spins directly, 
one would expect to observe their signatures in the zero-field resistance when  
Pr$_{1.29}$La$_{0.7}$Ce$_{0.01}$CuO$_4$ and Pr$_{1.85}$Ce$_{0.15}$CuO$_4$ 
order antiferromagnetically \cite{lavrov,fournier}, and when different 
noncollinear spin phase transitions occur in Nd$_{1.975}$Ce$_{0.025}$CuO$_4$ (Fig. 2). However, AF order
appears to have no observable effects on zero-field resistance. Instead, the maximum of $\Delta R_c/R_c(0)$ in Nd$_{1.975}$Ce$_{0.025}$CuO$_4$
shows a $1/T$ temperature dependence [inset of Fig. 8(b)], very similar to the $1/T$ temperature dependence of
the Nd moment in various Nd-containing Nd$_{2-x}$Ce$_{x}$CuO$_4$ 
compounds \cite{skanthakumarjap,skanthakumarprb,lynn,masato,kang}. This strongly suggests that the 
observed MR effects in Nd$_{1.975}$Ce$_{0.025}$CuO$_4$ and other electron-doped materials 
are somehow related to the rare earth (Nd,Pr) moments and/or Nd(Pr)-Cu coupling.
This picture may also explain why, when the dominant spin-spin interactions are from Cu-Cu with negligible 
Nd moments in the type-I collinear phase ($T> 68$ K), the weak MR data are essentially temperature
independent and collapse onto a single curve [Fig. 8(b)].

To compare our results with hole-doped materials, 
we note that large anisotropic MR effects have already been reported for lightly doped 
La$_{2-x}$Sr$_{x}$CuO$_4$ \cite{ando03} and YBa$_2$Cu$_3$O$_{6+x}$ \cite{ando,cimpoiasu}.
These results have been interpreted as due to the influence of an applied magnetic field on stripes \cite{ando03,ando,kivelson}, spin-orbital coupling \cite{moskvin}, 
redistribution of magneto-elastic antiferromagnetic domains \cite{gomonaj}, or 
canted AF spin structures \cite{juricic}. At present, there is no consensus
on a microscopic picture for the MR effects in hole-doped copper oxides and
more work is needed to test the predictions of different models.  However, regardless of the details for each 
model, what is clear is that transport properties of electron or hole doped copper oxides are closed related
to the AF order in these materials.

\section{Conclusions}
In summary, we have shown that spin-flop transition from noncollinear to collinear
state in a lightly electron-doped copper oxide affects both the in-plane and out-of-plane 
MR. The application of a in-plane magnetic field can induce $c$-axis spin disorder 
and hysteresis in the AF phase transitions.  By comparing neutron scattering results 
of Nd$_2$CuO$_4$ with  
the MR effects in
Nd$_{1.975}$Ce$_{0.025}$CuO$_4$, we show that the transport 
properties of these materials are very sensitive
to the  subtle changes in the spin structures. Our results thus provide further evidence 
for the existence of a strong spin-charge coupling in both electron and hole doped copper oxides.

\begin{acknowledgments}
We would like thank Yoichi Ando and Hai-hu Wen for helpful discussions.
This work is supported by the U. S. NSF DMR-0139882, DOE No. DE-AC05-00OR22725 
with UT/Battelle, LLC., and
by NSF of China under contract No. 10128409.

\end{acknowledgments}


\end{document}